\documentclass[aps,prl,twocolumn,floatfix,nofootinbib,showpacs,superscriptaddress]{revtex4-1}
\usepackage{amsmath,amsfonts,amssymb,bm}
\usepackage{graphicx}
\usepackage{subfigure}
\usepackage{color}
\definecolor{purple}{rgb}{0.5,0,0.5}
\definecolor{blue}{rgb}{0.0,0,0.9}
\usepackage[colorlinks=true, pdfstartview=FitV, linkcolor=purple, citecolor= purple, urlcolor=blue]{hyperref}

\begin{document}
\title{Temperature Effect on Shear and Bulk Viscosities of QCD Matter}
 \author{Fei Gao }
\affiliation{Department of Physics and State Key Laboratory of Nuclear Physics and Technology,
Peking University, Beijing 100871, China}
\affiliation{Collaborative Innovation Center of Quantum Matter, Beijing 100871, China}

\author{Yu-xin Liu }
\affiliation{Department of Physics and State Key Laboratory of Nuclear Physics and Technology,
Peking University, Beijing 100871, China}
\affiliation{Collaborative Innovation Center of Quantum Matter, Beijing 100871, China}
\affiliation{Center for High Energy Physics, Peking University, Beijing 100871, China}

\date{\today}

\begin{abstract}
We investigate the temperature dependence of the shear and bulk viscosities and their ratios to the entropy density via a continuum QCD approach.
We calculate the pion mass and decay constant in the framework of Dyson-Schwinger equations of QCD and the pion thermal width by combining with Roy equations.
We obtain then the variation behaviors of the viscosities, especially a novel feature of the bulk viscosity, with respect to temperature.
\end{abstract}


\maketitle

\noindent{\emph{Introduction:}}---It has been well known that the shear and bulk viscosities ($\eta$, $\zeta$) and their ratios to the entropy density ($s$) are excellent signatures to identify the characteristic of the correlation between the composing particles of a matter (See, e.g., Ref.~\cite{McLerrin:2006PRL}).
Investigations relating the measured transverse momentum correlations, energy loss, elliptic flow, etc, in  relativistic heavy ion collisions (RHIC)  to $\eta/s$
({\it e.g.}, Refs.~\cite{Gavin:2006PRL,Lacey:200714PRL,Adare:2007PRL,Wang:2007PRL,Song:2011PRL}) yield $1 < 4\pi \eta/s < 3.75$, which is quite close to its quantum lower bound $1/4\pi$ (KSS bound)~\cite{Kovtun:2005PRL} and indicate that the matter created by the RHIC experiments is in the strongly-coupled quark gluon plasma (sQGP) state.
Due to the sensitivity of the observables to the temperature and chemical potential ($T$, $\mu$) or the collision energy ($\sqrt{s_{NN}^{}}$) dependence of the $\eta /s$ (e.g., Refs.~\cite{McLerrin:2006PRL,Lacey:200714PRL,Gorenstein:2007PRC,Niemi:2011PRL}, the ratio may also be a signature of the critical end point (CEP) in the QCD phase diagram~\cite{Lacey:200714PRL,Ghosh:2015PRD}.
However, the temperature dependence of the $\eta/s$ is still controversial, some of them manifest the ``U"-like feature demonstrating  the crossover of the phase transition~\cite{McLerrin:2006PRL,Chen:2008PLB,Chakraborty:2011PRC,Bluhm:2011PRC,Plumari:2011PRD,Puglisi:2015PLB,Ghosh:2015PRD,Pawlowski:2015PRL,Huang:2015JHEP,Deb:2016PRD}, some others give only the decreasing behavior~\cite{Itakura:2008PRD,Demir:2009PRL,Greiner:2009PRL,Plumari:2010PLB,Pal:2010PLB,Fu:2013PRD,Marty:2013PRC,Ghosh:2014PRC,Kadam:2015NPA}  or the increasing feature~\cite{Nakamura:2005PRL,Qin:2014PLB,Ghosh:2016PRC1}.

The situation of the  bulk viscosity, $\zeta$, is more complicated. Generally, the $\zeta$ is recognized to be very small~\cite{Arnold:2006PRD,FernandezFraile:2009PRL,Dobado:2012PRD,Marty:2013PRC,Mitra:2013PRD,Ghosh:2016PRC2}, but  theoretical studies relating the $\zeta$ with the trace anomaly~\cite{Karsch:2008PLB,Meyer:2008PRL,Kharzeev:2008JHEP} obtain and explaining recent experiments requires~\cite{Ryu:2015PRL,Heinz:2016PRC} a large value.
Nevertheless, the relation connecting the bulk viscosity with the trace anomaly~\cite{Kharzeev:2008JHEP}
is very subtle since it requires particular ansatz for the spectral function~\cite{Moore:2008JHEP,Sasaki:2009PRC} and sophisticated technique to rule out the non-thermal contributions~\cite{Huebner:2008PRD,Arnold:2006PRD}.
Moreover the presently existing temperature dependence of the $\zeta/s$ has not reached a common idea either, {\it e.g.},  lattice QCD simulations give a concave function (diverging at the chiral critical temperature ($T_{c}^{\chi}$) and then decreasing~\cite{Karsch:2008PLB}, or taking maximum at temperature slightly above the $T_{c}^{\chi}$~\cite{Meyer:2008PRL}), holographic correspondence of the supersymmetric Yang-Mills theory shows a similar feature but the maximum appearing at lower temperature~\cite{Buchel:2014PLB} or at about $T_{c}^{\chi}$~\cite{Kampfer:2015PLB,Huang:2015JHEP}, chiral perturbation theory gives double peaks relation~\cite{FernandezFraile:2009PRL}. More contradictorily, some of hadron resonance model and Nambu--Jona-Lasinio (NJL) model calculations yield convex or concave function for $\zeta/s (T)$~\cite{Kadam:2015NPA,Ghosh:2016PRC1}, some others give a decreasing feature~\cite{Plumari:2010PLB,Marty:2013PRC,Hou:2014CPC,Deb:2016PRD,Ghosh:2016PRC2} or an increasing behavior~\cite{Greiner:2009PRL}; while quasiparticle model produces decreasing or concave function if the mass of the $\sigma$-meson take different values~\cite{Chakraborty:2011PRC}, linear sigma model yields convex or concave or more complicated behavior depending the masses of the pion and $\sigma$-meson~\cite{Dobado:2012PRD}. Besides, controversial results also exist for its $(1/3 - c_{s}^{2})$ (where $c_{s}^{2}$ is the sound speed squired) dependence (see, {\it e.g.}, Refs.~\cite{Benincasa:2006NPB,Arnold:2006PRD,Buchel:2008PLB,Klimek:2011PLB,Kampfer:2015PLB}).
It is then imperative to clarify the problems with sophisticated QCD approaches.

It has been known that the Dyson-Schwinger equations (DSEs) (see, e.g., Ref.~\cite{Roberts:DSE-BSE}) are almost uniquely a continuum QCD approach which includes both the confinement and the dynamical chiral symmetry breaking (DCSB) simultaneously~\cite{McLerran:2007NPA},
and are successful in describing  QCD phase transitions and hadron properties (see, e.g., Refs.~\cite{Roberts:DSE-BSE,QCDPT-DSE}).
We then, in this Letter, take the DSE method to shed light on the issues.

\medskip

\noindent{\emph{Framework:}}
---Strong interaction matter (QCD matter) can be considered as a system consisting of quarks and antiquarks (most of them are confined to form pions, the lowest mass hadrons, at low temperature).
The transport coefficients of the system can be obtained in the kinetic theory approach~\cite{Chakraborty:2011PRC,Ghosh:2014PRC,FernandezFraile:2009PRL,Ghosh:2016PRC2}.
The standard expressions are£º
\begin{eqnarray}
\eta & = & \frac{1}{10\pi^{2} T} \! \int_{0}^{\infty} \!\! \frac{d|\vec{k}| |\vec{k}|^6}{(\omega_{|\vec{k}|}^{\pi})^{2} \Gamma_{\pi}^{\textrm{TW}}(|\vec{k}|)} {n_{|\vec{k}|}^{}}(\omega_{|\vec{k}|}^{\pi}) \! \left\{1 \! + \! n_{|\vec{k}|}^{}(\omega_{|\vec{k}|}^{\pi})\right\}, \quad \notag  \\
\zeta &= &\frac{3}{T} \! \int \! \frac{d^{3}\vec{k}}{(2\pi)^{3}}\frac{1}{(\omega_{|\vec{k}|}^{\pi})^{2} \Gamma_{\pi}^{\footnotesize{\textrm{TW}}}(|\vec{k}|)}n_{|\vec{k}|}^{}(\omega_{|\vec{k}|}^{\pi})
\! \left\{1\! + \! n_{|\vec{k}|}^{}(\omega_{|\vec{k}|}^{\pi})\right\} \quad \notag  \\
 &&\times\left\{ \Big{(}\frac{1}{3} - c_{s}^{2} \Big{)} |\vec{k}|^{2} - c_{s}^{2} m_{\pi}^{2} \right\} ,
\end{eqnarray}
where $n_{|\vec{k}|}^{}(\omega_{|\vec{k}|}^{\pi})=1/(e^{\omega_{|\vec{k}|}^{\pi}/T}-1)$ is the Bose-Einstein distribution with $\omega_{|\vec{k}|}^{\pi}=\sqrt{{|\vec{k}|}^{2}+m_{\pi}^{}}$, $m_{\pi}^{}$ is the pion mass, and $\Gamma_{\pi}^{\footnotesize{\textrm{TW}}}({|\vec{k}|})$ is the pion thermal width.
$c_{s}^{2}$ is the sound speed squared defined as $c_{s}^{2}=\frac{\partial P}{\partial \varepsilon}$ with $P$ the pressure and $\varepsilon$ the energy density.
%
%

\medskip

\noindent{\emph{Pion Property:}}---It is apparent that, to investigate the property of the viscosities,
one needs at first the thermal property of pion.
The property of the pion in vacuum has been attentively investigated in the framework of DSEs of QCD.
Within the bare vertex truncation for quark propagator DSE, it is accurate for light-flavor groundstate vector- and pseudoscalar-mesons~\cite{Maris:19979PRC,Chang:200912,Roberts:DSE-BSE}.
It has also been proved that in chiral limit, the massless pion can be connected with the DCSB with formula: $f_{\pi}^{} E(p)=B_{0}^{}(p)$, where $f_{\pi}^{}$ is the pion decay constant,
$E(p)$ the dress function for component $\gamma_{5}^{}$ of pion's Bethe-Salpeter amplitude $\Gamma_{\pi}^{}(p)$, $B_{0}^{}(p)$ the scalar part of the inverse of quark propagator in chiral limit~\cite{Maris:1998PLB}.
In case of finite temperature and beyond the chiral limit, one may take the similar scheme to describe the meson properties,
that is, $\Gamma_{\pi}^{}(p) \propto i\gamma_{5}^{} B(p)$ with $B(p)$ the scalar part of the inverse quark propagator at respective current quark mass and $p = (\vec{p}, \omega_{n} )$, if proper ultraviolet behavior is taken into account.
After considering the right ultraviolet feature, we get the explicit relation as:
\begin{eqnarray}
\Gamma_{\pi}^{}(p) & = & i\gamma_{5}^{} B_{r} (p)/{f_{\pi}^{}} \, , \label{eq:pion0} \\[1mm]
{\textrm{with}} \qquad B_{r} & = & B(p) - m \frac{\partial B(p)}{\partial m} \, , \qquad \qquad \notag
\end{eqnarray}
where $m$ is the current mass of the light quark.
The pion mass and decay constant can be formulated as similar as those in Ref.~\cite{Bender:1998PLB}:
\begin{eqnarray}
f_{\pi}^{} N_{\pi}^{} & = & 4 N_{c} \int_{p} B_{r} \Big{\{} \sigma_{A}^{} \sigma_{B}^{} +\frac{2}{3}|\vec{p}|^{2}(\sigma_{A}^{\prime} \sigma_{B}^{} - \sigma_{A}^{} \sigma_{B}^{\prime}) \Big{\}} \, ,  \label{eq:pion1}  \\
m_{\pi}^{2} f_{\pi}^{2} & = & 8 N_{c} \int_{p} \frac{m B_{r}^{}}{\omega_{n}^{2} C^{2}
+ |\vec{p}|^{2} A^{2} + B^{2}} \, , \label{eq:pion2}\\
N_{\pi}^{2} & = & 2 N_{c} \! \! \int_{p} \! B_{r} \Big{\{} \sigma_{A}^{2} -2(\omega_{n}^{2} \sigma_{C}^{} \sigma_{C}^{\prime} +|\vec{p}|^{2}\sigma_{A}^{} \sigma_{A}^{\prime}
+\sigma_{B}^{} \sigma_{B}^{\prime}) \notag\\
&& -\frac{4}{3}|\vec{p}|^{2} \big{[} \omega_{n}^{2} [\sigma_{C}^{}\sigma_{C}^{\prime\prime} -(\sigma_{C}^{\prime})^2 ]
+|\vec{p}|^{2} [\sigma_{A}^{} \sigma_{A}^{\prime\prime} - (\sigma_{A}^{\prime})^{2} ] \notag\\
&&+\sigma_{B}^{} \sigma_{B}^{\prime\prime} - (\sigma_{B}^{\prime})^2 \big{]} \Big{\}} \, , \label{eq:pion3}
\end{eqnarray}
where $N_{\pi}^{}$ is the normalisation factor of Bethe-Salpeter amplitudes. $\sigma_{F}^{}$ and $F$ ($F=$ $A$ ,$B$, $C$) is defined in the quark propagator as
$$S(p) = (i\vec{\gamma}\cdot \vec{p}A + \gamma_{4}^{} \omega_{n}^{}\cdot C+B)^{-1}
= i\vec{\gamma}\cdot \vec{p}\sigma_{A}^{} + \gamma_{4}^{} \omega_{n}^{} \cdot \sigma_{C} +\sigma_{B} , $$
and
$\sigma_{F}^{\prime} =\partial{\sigma_{F}^{}}/(\partial |\vec{p}|^{2})$,
which can be determined with the DSE:
\begin{eqnarray}
\nonumber
S(\vec{p}, i\omega_{n}^{})^{-1} & = & Z_{2}^{A} i \vec{\gamma} \cdot \vec{p} + Z_{2}^{} i \gamma_{4}^{}
\omega_{n}^{} + Z_{4}^{} m \\
& &  + T \sum_{l} \frac{4}{3} \int \frac{d^3 q}{(2\pi)^3}
g^{2} D_{\mu\nu}(\vec{p} \! - \! \vec{q},\Omega_{nl})\notag    \\
& & \; \; \times \big{(} Z_{1}^{A}\gamma_{\mu}^T + Z_{1} \gamma_{\mu}^{L} \big{)}
S(\vec{q}, i\omega_{l}^{}) \Gamma_{\nu} \, , \label{eq:gapeq}
\end{eqnarray}
where $\Gamma_{\nu} = \Gamma_{\nu} (\vec{q}, \omega_{l}^{}, \vec{p},
\omega_{n}^{} )$ is the quark-gluon interaction vertex, $Z_{2}^{A}$, $Z_{2}$, $Z_{4}$, $Z_{1}^{A}$, $Z_{1}$ is the respective renormalisation factor,
$\Omega_{nl}=\omega_{n}^{} - \omega_{l}^{}$ with $\omega_{n}^{} = (2n +1)\pi T$ being the quark's Matsubara frequency, and $\gamma^{T,L}_\mu=P_{\mu\nu}^{T,L} \gamma_{\nu}^{}$ with $P_{\mu\nu}^{T,L}$ the transverse and longitudinal projection operators, respectively.
$g^{2} D_{\mu\nu}(\vec{k}, \Omega_{nl}) $ is the interaction kernel together with the dressed-gluon propagator, which has generally the form
\begin{eqnarray}
g^{2} D_{\mu\nu}(\vec{k}, \Omega_{nl}) & = & P_{\mu\nu}^{T} D_{T}(\vec{k}\,^{2}, \Omega_{nl}^{2}, m_{g}^{2}) \notag \\
& & + P_{\mu\nu}^{L} D_{L}(\vec{k}\,^{2}, \Omega_{nl}^{2}, m_{g}^{2})\, .
\end{eqnarray}
Extending the infrared constant model proposed in Ref.~\cite{Qin:2011PRC}, we can have
\begin{eqnarray}
\label{gluon}
\mathcal{D}_{L,T}(\vec{k}\,^{2}, \Omega_{nl}^{2}, m_{g}^{2}) & = & 8{\pi^{2}} {\mathcal{D}}
\frac{1}{\omega_{L,T}^{4}} e^{-{s_{\Omega}^{}}/\omega_{L,T}^{2}} \notag\\
& & + \frac{8{\pi^{2}} {\gamma_{m}}{\cal F}(s_{\Omega}^{}) }{{\ln}[ \tau \! + \! (1 \! + \!
{s_{\Omega}^{}}/{\Lambda_{\text{QCD}}^{2}} ) ^{2} ] } \, ,
\end{eqnarray}
with ${\cal F}(s_{\Omega}^{}) = (1-\exp(-s_{\Omega}^{}/4 m_{t}^{2}))/{s_{\Omega}^{}}$,  $s_{\Omega}^{} = \Omega^2 + \vec{k}^{2}+m_{g}^{2}$ with $m_{g}^{}$ the thermal mass of gluon, $\gamma_{m}^{} = 12/25$,  $\tau=e^{2}-1$, $m_{t}^{} =0.5\,$GeV, $\Lambda^{}_{\text{QCD}}=0.234\,$GeV.
It has been known that, taking the infrared constant gluon propagator~\cite{Qin:2011PRC} (i.e., $\omega_{L} = \omega_{T} = \omega$) with
parameter(s) ${\mathcal{D}}=1.024\,\textrm{GeV}^{2}$ and $\omega=0.5\,$GeV, one can describe some mesons' properties at $T=0$ quite well.
However, if we compare the kernel with the results given in recent Lattice QCD simulations and other modern DSE calculations (see, e.g., Refs.~\cite{Silva:2014PRD,Fischer:2014PRD}),
we find that extending the infrared constant model directly to finite temperature can not describe the difference between the temperature dependence of the longitudinal part and that of the transverse part of the gluon propagator. We then correct the constant $\omega$ with temperature dependence as
\begin{equation}
\omega_{L}^{}=0.5-T \, , \qquad  \omega_{T}^{}=0.5 +1.6T\, .
\end{equation}
It should also be mentioned that such a correction is only appropriate for the states before the phase transition.
After the transition from hadrons to quark-gluon states,
%
%
to include the temperature screening effect explicitly we take a damping factor for the ${\mathcal{D}}$:  $m_{g}^{2} = 0$ for the ${\mathcal{D}}_{T} $ and $m_{g}^{2}=16 T^{2} /5 $ for the ${\mathcal{D}}_{L} $ according to perturbative QCD calculations~\cite{Thoma:1998NPA,Haque:2013PRD}.

\begin{figure}[htb]
\centerline{\includegraphics[width=0.440\textwidth]{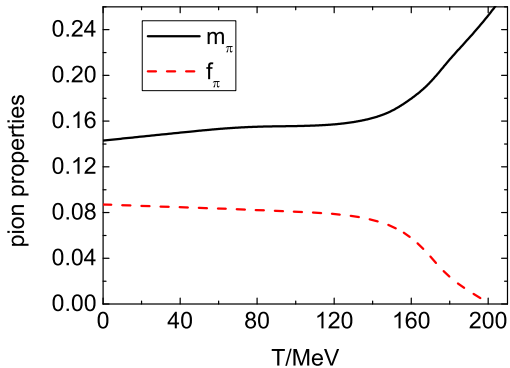}}
\caption{(color online) Calculated temperature dependence of the pion's mass (\emph{solid}) and decay constant (\emph{dashed})} \label{fig:pion}
\end{figure}
We then take the bare vertex approximation $\Gamma_{\nu} = \gamma_{\nu}^{}$ which has been known to be appropriate to describe the pion properties and parameters $m=3.4\,$MeV, $D=1.024\,\textrm{GeV}^{2}$ and $\omega=0.5\,$GeV to determine the pion mass and decay constant with Eqs.~(\ref{eq:pion0})-(\ref{eq:pion3}).
The obtained results are shown in Fig.~\ref{fig:pion}.
It is evident that the $m_{\pi}^{}=143\,$MeV and $f_{\pi}^{}=87\,$MeV at $T=0$
agree with empirical data very well.
The decreasing behavior of the $f_{\pi}^{}(T)$ and the increasing feature of the $m_{\pi}^{}(T)$ are also consistent with previous results.

To evaluate the pion thermal width, one usually resort to the properties of the resonant states in   $\pi$--$\pi$ scattering, It has been well known that $\sigma$- and $\rho$-mesons are, respectively, the $S$- and $P$-wave resonant state of the scattering.
The width $\Gamma_{\pi}^{\footnotesize{\textrm{TW}}}$ can then be
formulated as~\cite{Ghosh:2014PRC}:
\begin{equation}
\label{eq:thermalwidth}
\Gamma_{\pi}^{\footnotesize{\textrm{TW}}}  =  \sum_{i=\rho,\sigma}\frac{1}{N_{i}}\int^{(m^{+}_{i})^2}_{(m^{-}_{i})^2}dM^2
\rho_{i}^{}(M) \Gamma^{}_{\pi\pi,i}\, ,
\end{equation}
with
$$
 \Gamma^{}_{\pi\pi,i} = \frac{1}{16\pi m_{\pi}^{} |\vec{k}|} \int^{\omega_{-}^{}}_{\omega_{+}^{}}\! d\omega L_{i}(\omega)  \big{[} n(\omega) - n(\omega^{\pi}_{|\vec{k}|}+\omega) \big{]} \, ,
$$
where $m^{\pm}_{i} = M_{i}^{} \pm\Gamma_{i}^{}$ with $M_{i}^{}$, $\Gamma_{i}^{}$ being the mass, the width of the resonance $i$, respectively.
$\omega^{\pm} =R^{2}(-\omega^{\pi}_{|\vec{k}|}\pm|\vec{k}|(1 - 1/R^{4})^{1/2})$
with $R^{2} = 1- \frac{M_{i}^{2}}{2m^{2}_{\pi}}$.
%
%
With the narrow-width approximation,
the resonance width $\Gamma_{i}^{}$ can be fixed by the Breit-Wigner formula
\begin{equation}
\rho_{i}^{}(M) = \frac{1}{\pi} \textrm{Im}\! \Big{[} \frac{-1}{M^{2} - M_{i}^{2} + i\Gamma_{i} M_{i}^{}} \Big{]} . \end{equation}
In one-loop calculation, $L_{\sigma}^{}(\omega)=\frac{-g_{\sigma}^{2} M_{\sigma}^{2}}{4}$
and $L_{\rho}^{}(\omega)=\frac{-g_{\rho}^{2}}{M_{\rho}^{2}} \big{[} 2m_{\pi}^{2}(m_{\pi}^{2} - M_{\rho}^{2})
-2(-\omega_{|\vec{k}|}^{\pi} \omega M_{\rho}^{2}  + m_{\pi}^{4}) \big{]}$,
with parameters $g_{\rho}^{}=6$ and $g_{\sigma}^{}=6.85$.

It is very hard to compute the mass and width of the $\rho$- and $\sigma$-resonances in this scenario
directly in the DSE scheme (in fact, it is impossible for the width of mesons at present stage).
We then appeal to the Roy equations.
In the framework of Roy equations~\cite{Roy:1971PLB,Leutwyler:2001PRt}, the partial wave amplitude $t_{l}^{I}(s)$ can be connected with the  scattering lengths.
The amplitudes contain mass poles of the resonance on the second sheet of the $s$-plane,
if and only if the S matrix, $S(s)=1-2\sqrt{4 m_{\pi}^{2}/s - 1}t(s)$, has a zero on the physical sheet~\cite{Caprini:2006PRL}.
The position of the zero is $s=M_{i} + \frac{i}{2}\Gamma_{i}$ with $i=\sigma$ for $t_{0}^{0}(s)$
and $i=\rho$ for $t_{1}^{1}(s)$, respectively.
The Roy equations are a set of simultaneously coupled integral equations of various channels.
Solving the Roy equations including the relevant ones $t_{0}^{0}$, $t_{0}^{2}$ and $t_{1}^{1}$ (for their explicit expresses, see Ref.~\cite{Pennington:1973PRD}) with the scattering lengths being set as the Weinberg's formulae~\cite{Weinberg:1966PRL}:
$a_{0}^{0} = \frac{7 m_{\pi}^{2}}{32\pi f_{\pi}^{2}}$ and $a_{0}^{2} = -\frac{m_{\pi}^{2}}{16\pi f_{\pi}^{2}}$,
we get the temperature dependence of the mass and the width of the $\sigma$- and $\rho$-resonances
with the above obtained pion mass and decay constant at finite temperature as input.
The obtained results of the masses and their widths at zero temperature ($m_{\sigma,0}= 427\,$MeV, $\Gamma_{\sigma,0}= 624\,$MeV, $m_{\rho,0}= 770\,$MeV, $\Gamma_{\rho,0}= 134\,$MeV) agree with experimental data~\cite{PDG} very well.
As temperature increases, the widthes of the $\rho$- and $\sigma$-mesons decrease,
and the masses of the mesons decrease at low temperature and turn to increase after reaching their respective  minimum.
At $T \approx 183\,$MeV, $M_{\sigma}^{} = M_{\rho}^{}= 2 M_{\pi}^{} = 4M_{q}^{}$. It manifests that the mesons get melted to quarks, i.e., the deconfinement takes place.

%

\begin{figure}[htb]
\centerline{\includegraphics[width=0.480\textwidth]{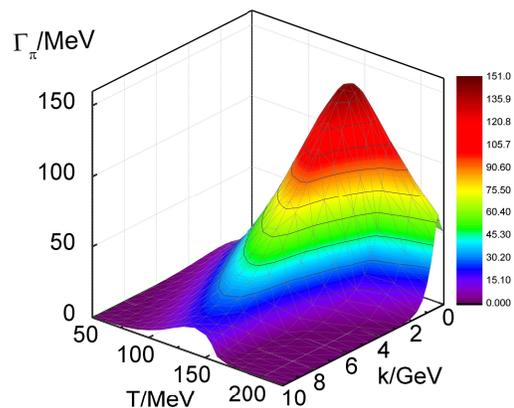}}
\caption{(color online) Calculated temperature and momentum dependence of the pion thermal width} \label{fig:thermalwidth}
\end{figure}

With the above obtained resonance properties of the $\pi$-$\pi$ scattering and Eq.~(\ref{eq:thermalwidth}), we get further the thermal width of pion. The obtained momentum and temperature dependence of the  width is displayed in Fig~\ref{fig:thermalwidth}.

\medskip

\begin{figure}[htb]
\centerline{\includegraphics[width=0.480\textwidth]{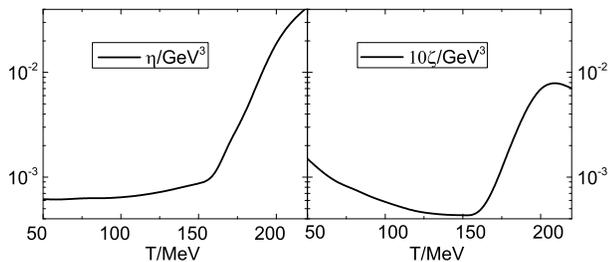}}
\caption{Calculated temperature dependence of the shear viscosity $\eta$ (\emph{left panel}) and the bulk viscosity $\zeta$ (\emph{right panel}) } \label{fig:transT}
\end{figure}

\noindent{\emph{Numerical Results of the Transport Coefficients:}}---After obtaining the pion mass and its thermal width $\Gamma_{\pi}^{\footnotesize{\textrm{TW}}}$, we calculate the shear and bulk viscosities of the matter. The obtained temperature dependence of the viscosities are shown in Fig.~\ref{fig:transT}.
The figure manifests obviously that the shear viscosity ascends slightly at low temperature,
and increases rapidly at the temperature around the chiral phase transition ($\chi$PT, in fact, a crossover).
Meanwhile, at low temperature, the bulk viscosity shows a decreasing behaviour till the $\chi$PT takes place, and the value of bulk viscosity is quite small in this region (about $10\%$ of the shear viscosity).
However, in the temperature region for the $\chi$PT to happen,
the bulk viscosity demonstrates an obvious bump behaviour,
and the value becomes large which can give a nonnegligible effect on
the experiment signal of QCD matter.

\begin{figure}[htb]
\centerline{\includegraphics[width=0.480\textwidth]{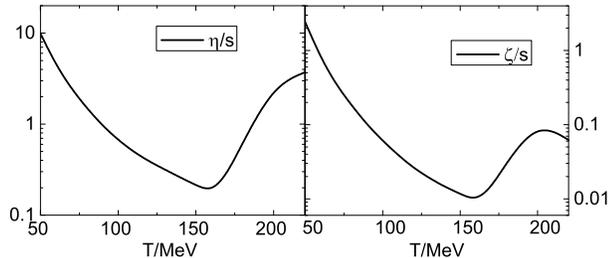}}
\caption{Calculated temperature dependence of the $\eta/s$ (\emph{left panel}) and the  $\zeta/s$ (\emph{right panel}) } \label{fig:trans}
\end{figure}

We compute further the ratios of $\eta/s$ and $\zeta/s$  with the entropy density expressed as
$$s=\frac{\partial P}{\partial T} = \frac{3}{T} \int\frac{d^{3} \vec{k}}{(2\pi)^{3}} \Big{(} \omega_{|\vec{k}|}^{\pi} + \frac{\vec{k}^{2}}{3\omega_{|\vec{k}|}^{\pi}} \Big{)} n_{|\vec{k}|}(\omega_{|\vec{k}|}^{\pi})\, . $$

The obtained temperature dependence of the ratios $\eta/s$ and $\zeta/s$ are illustrated in Fig.~\ref{fig:trans}.
One can observe obvious minima for both the $\eta/s$ and the $\zeta/s$ around the pseudo-critical temperature $T_{c}^{\chi}$ ($\sim \! (154 \pm 9)\,$MeV~\cite{Tc-lattice}). More concretely, the temperature for the $\eta/s$ to have its minimum is $T=158.8\,$MeV, while for $\zeta/s$ it is $T=155.8\,$MeV.
The minimal value of the $\eta/s$ is $0.202$, which is about $\frac{2.5}{4\pi}$.
This result indicates that, if we take into account the temperature effect on the mesons involved carefully, the ratio $\eta/s$ of the strong interaction matter won't break the KSS bound.
Nevertheless, such a small value still shows the  strong-coupled property of the matters
near the $\chi$PT temperature.
Meanwhile, we should notice that in the low temperature region,
the ratio $\eta/s$ experiences a dramatic decrease as the temperature ascends,
and there would be  large error if extracting the shear viscosity of QGP
with a constant shear viscosity for hadronic phase.
As for the $\zeta/s$,
%
%
it holds a minimum with value about $0.01$ and, specifically, a maximum about $0.11$ at $T \approx 200\,$MeV.
Although the value is still much smaller than the $\eta /s$ at same temperature (about $10\%$),
it is not negligible.
Such a bump behaviour indicates that the proposal that relates the bulk viscosity to the trace anomaly is convincible and it may be a demonstration of the phase transition.
It is also notable that,
our results leads to $\zeta/\eta \approx (0.82 \sim 1.16) (1/3-c_{s}^{2})$ and
$\zeta/\eta \approx (0.73 \sim 1.66) 15(1/3 - c_{s}^{2})^{2} $ around the temperature $T=150 \sim 200\,$MeV.
It seems that our result near the phase transition region is slightly in favour of the estimation from the strong-coupled non-conformal gauge/gravity dual theory~\cite{Benincasa:2006NPB}
than of the leading-order perturbative QCD result~\cite{Arnold:2006PRD}.

\medskip

\noindent{\emph{Summary:}}---
In the framework of Dyson-Schwinger equations of QCD and the Roy equations of the pion-pion scattering, we studied the shear viscosity and the bulk viscosity and their temperature dependence of QCD matter in this Letter.
We compute firstly the temperature dependence of the pion mass and decay constant,
then the properties of $\sigma$- and $\rho$-mesons in the resonance of $\pi$-$\pi$ scattering, and obtain the temperature and momentum of the thermal width of pion.  %
We gain eventually the temperature dependence of the shear viscosity and bulk viscosity and their ratios to entropy density of the QCD matter, more specifically, the pion gas.
Our results indicate that, as the temperature ascends, the $\eta/s$ behaves a concave function
and the $\zeta/s$ a concavo-convex function.
Both of the ratios hold their minimum around the pseudo-critical temperature $T_{c}^{\chi}$ of the chiral phase transition (crossover), with value $0.202$, $0.01$, respectively.
%
%
Meanwhile, the obtained bulk viscosity is generally much smaller.
The bump with maximum value about $0.11$ of the $\zeta/s$ in the phase transition region manifests that the matter at temperature around $1.3\,T_{c}^{\chi}$ is still in the sQGP state and supports the proposal that relates the bulk viscosity to the trace anomaly.

\medskip

The work was supported by the National Natural Science Foundation of China under Contract No. 11435001; the National Key Basic Research Program of China under Contracts No.~G2013CB834400
and No.~2015CB856900.

\end{document}